\documentclass[twocolumn,showpacs,preprintnumbers,amsmath,amssymb,superscriptaddress]{revtex4}

\usepackage{graphicx}%
\usepackage{dcolumn}%
\usepackage{bm}%
\usepackage{color}
 \usepackage{subfigure}

\begin{document}

\title{Dissipative phase solitons in semiconductor lasers}%

\author{F. Gustave}
\affiliation{Universit\'e de Nice CNRS, Institut Non Lin\'eaire de Nice, 1361 route des lucioles 06560 Valbonne, France}
\author{L. Columbo}
\affiliation{Dipartimento Interateneo di Fisica, Universit$\grave{a}$ degli Studi e Politecnico di Bari, Via Amendola 173, 70126 Bari, Italy}%
\author{G. Tissoni}
\affiliation{Universit\'e de Nice CNRS, Institut Non Lin\'eaire de Nice, 1361 route des lucioles 06560 Valbonne, France}
\author{M. Brambilla}
\affiliation{Dipartimento Interateneo di Fisica, Universit$\grave{a}$ degli Studi e Politecnico di Bari, Via Amendola 173, 70126 Bari, Italy}%
\author{F. Prati}
\affiliation{Dipartimento di Scienza e Alta Tecnologia, Universit$\grave{a}$ dell'Insubria, Via Valleggio 11, 22100 Como, Italy}%
\author{B. Kelleher}
\affiliation{Centre for Advanced Photonics and Process Analysis \& Department of Physical Sciences, Cork Institute of Technology, Cork, Ireland}
\author{B. Tykalewicz}
\affiliation{Centre for Advanced Photonics and Process Analysis \& Department of Physical Sciences, Cork Institute of Technology, Cork, Ireland}
\author{S. Barland}
\email{stephane.barland@inln.cnrs.fr}
\affiliation{Universit\'e de Nice CNRS, Institut Non Lin\'eaire de Nice, 1361 route des lucioles 06560 Valbonne, France}

\date{\today}%

\begin{abstract}
We experimentally demonstrate the existence of non dispersive solitary waves associated with a 2$\pi$ phase rotation in a strongly multimode ring semiconductor laser with coherent forcing. Similarly to Bloch domain walls, such structures host a chiral charge. The numerical simulations based on a set of effective Maxwell-Bloch equations support the experimental evidence that only one sign of chiral charge is stable, which strongly affects the motion of the phase solitons. Furthermore, the reduction of the model to a modified Ginzburg Landau equation with forcing demonstrates the generality of these phenomena and exposes the impact of the lack of parity symmetry in propagative optical systems. 
\end{abstract}

\pacs{Valid PACS appear here}%
\maketitle
Dissipative solitary waves have been observed as self-localized optical wave packets along the direction of propagation in many out of equilibrium and nonlinear optical systems.  In spite of their huge variety, many of the observations reported so far can be cast in two main categories depending on the presence or absence of coherent energy input, \textit{ie} the (lack of) phase symmetry of the system \cite{nphoton.2010.1}. In systems with phase symmetry, mode-locked laser pulses have been analyzed as dissipative solitons of the cubic-quintic Ginzburg Landau equation \cite{Grelu2012}. Their optical phase can wander in the course of time due to the neutral mode created upon the formation of a coherent wave. On the contrary, dissipative solitons in forced systems \cite{temporalcs,herr2014temporal} have been analyzed in the framework of the Lugiato-Lefever equation \cite{PhysRevLett.58.2209}, which includes a coherent forcing term acting as a phase reference to which solitons will lock. In both cases, the use of paradigmatic equations in addition to system-specific models has allowed to formally connect these optical solitary waves to localized states as they appear in fluid dynamics, plant ecology, granular media or reaction-diffusion systems \cite{akhmediev,akhmediev2008dissipative,descalzi2011localized,tlidi2014localized}. In fact, optical dissipative solitons can often be explained as perturbed solitons of the nonlinear Schr\"odinger equation (in the weak dissipation limit \cite{Nozaki1984383,malomed1987evolution,fauve90}) or as locked fronts (in strongly dissipative systems \cite{coullet00,rosanov_spatial_hysteresis}). 

In this Letter, we report on dissipative solitons which fundamentally consist of self-confined $2\pi$ phase rotations embedded in a homogeneously phase locked background. These \lq\lq phase solitons" are generic features of spatially extended oscillatory media under nearly resonant forcing \cite{coullet1992strong,chate199917} and result from the mismatch between the natural periodicity and the forcing. Here, the mismatch between the free running laser and the external forcing frequencies leads to the formation of phase kinks as result of a commensurate-incommensurate transition \cite{PhysRevLett.56.724,315223}. This connects our observations with the kink solutions observed in many physical systems described by the Frenkel-Kontorova model \cite{braun2004frenkel}, such as fluxons in Josephson arrays \cite{pfeiffer2006observation}, local deformations in DNA chain \cite{yakushevich2007influence}, or excitable waves in chemical and biological systems \cite{kuramoto2003chemical,coulletexcitwaves}. In nonvariational systems, chirality acquired through a non equilibrium Ising-Bloch transition implies the motion of domain walls \cite{coullet1990breaking,michaelis2001universal,gomila2015theory}. A link can thus be established with our observations of solitons that indeed move in the reference frame
of the propagating carrier wave (at light speed). Due to the propagative nature of the experiment and the non instantaneous semiconductor medium the system lacks parity symmetry, which impacts the chiral charge and therefore the motion of the phase solitons in this reference frame.

\begin{figure}[t]
\includegraphics[width=3.5cm]{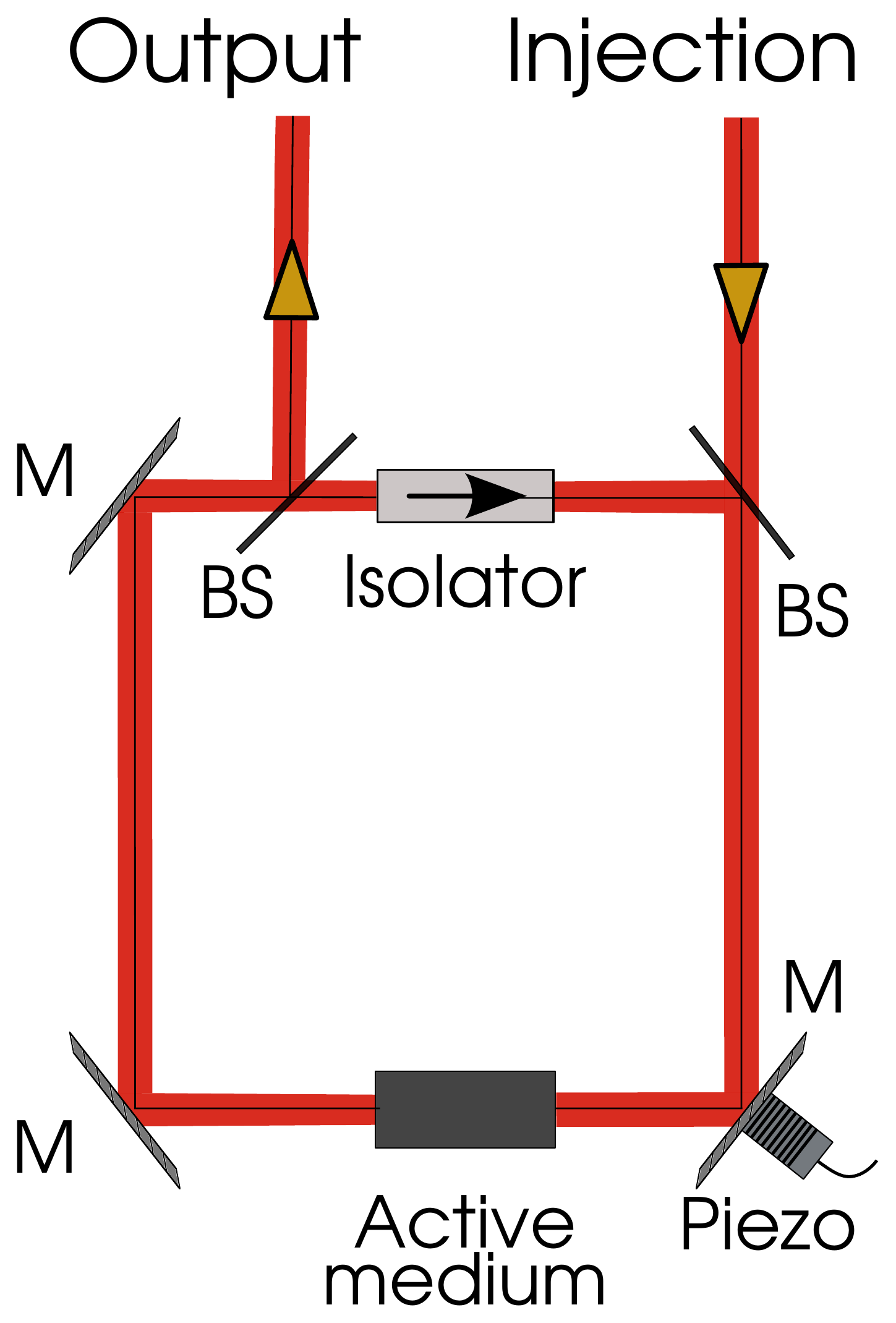}
\raisebox{-.2cm}
{\includegraphics[width=3.5cm]{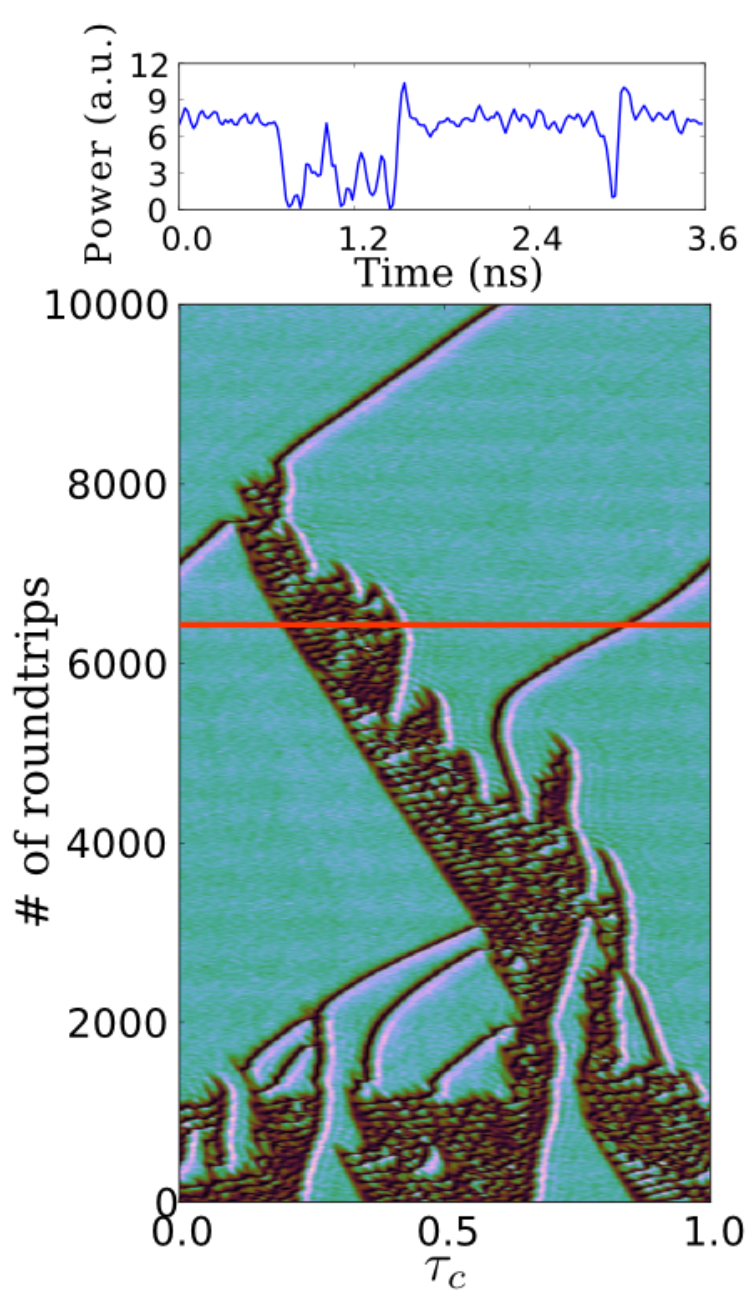}}
\caption{Left: Scheme of the experimental setup. Right: Color coded \protect{\cite{green2011colour,greenfootnote}} emitted intensity in a space-time representation. The horizontal axis is a fast time scale normalized to the round-trip time of the ring cavity $\tau$ (close to 3.6~ns) and the vertical axis is a slow time scale, in units of $\tau$.  \label{fig:setup}}
\end{figure}

The experimental set--up is based on a very large (length $\Lambda \simeq 1$~m) ring laser featuring a 4~mm-long ($l$) antireflection--coated semiconductor  medium inside a free space optical cavity (Fig.~\ref{fig:setup}, left) with low mirror transmissivity $T \approx 10\%$. The device operates in a regime where the field envelope in the single-mode regime would evolve on slower time scale than the active medium \cite{tierno:12}. In order to avoid transverse effects and directional competition \cite{6420863}, the optical cavity includes an aperture and an optical isolator.
The phase symmetry of the free running laser is broken by applying an external field provided by a grating tunable edge--emitting laser followed by an optical amplifier. The emitted intensity is acquired by a $9.5$~GHz photodetector coupled to a $12.5$~GHz real--time oscilloscope.\\
The laser is brought about 10\% above threshold and emits a superposition of many longitudinal modes. When sufficiently strong external forcing is applied, the system may lock in phase to the injection beam. In order to visualize the dynamics, we construct a co-moving space-time representation. We acquire very long time series (up to $10^7$ points) in real time and then split them into segments of equal length, corresponding to the time it takes for light to complete a full round-trip of the ring cavity. The segments are then stacked on top of each other, each of them showing the state of the system at a determined round-trip number. In this way, we obtain a spatiotemporal representation of the evolution of a 1-D system, the horizontal dimension showing the state of the system at a given moment and the vertical dimension its evolution over a discrete time measured in units of round-trips. Thanks to very well separated time scales (dynamics during one round-trip and evolution over the round-trips), the resulting space-time diagrams display very clearly the evolution of the system.

An example is shown on Fig.~\ref{fig:setup} (right). This regime is experimentally attained by choosing initially a very strong injected power (sufficient to  phase lock the ring laser) and setting the detuning to bring the system very close to unlocking. Then, upon progressive decrease of the injection power and adjustment of the detuning to remain close to the unlocking transition, chaotic regions spontaneously appear. Here, during the first 1000 round-trips, a few spatial regions are in a complex dynamical state with only small segments locked to the external forcing. Between round-trips 1000 and 3000 a few isolated structures emerge from the chaotic domains and drift towards the right, eventually hitting a larger chaotic domain which seemingly absorbs them. At rountrip 4000 a new solitary wave emerges from the chaotic domain and drifts first to the left and later to the right, going through the boundary on the right and emerging on the left about round-trip 7000. The time trace shown on top of right panel is a cut through the space-time diagram taken at round-trip 6300 (red line). We attribute the strong asymmetry between left and right to the lack of parity symmetry (also discussed below) in this propagative system with non-instantaneous medium.

\begin{figure}[t]
\includegraphics[width=4.5cm]{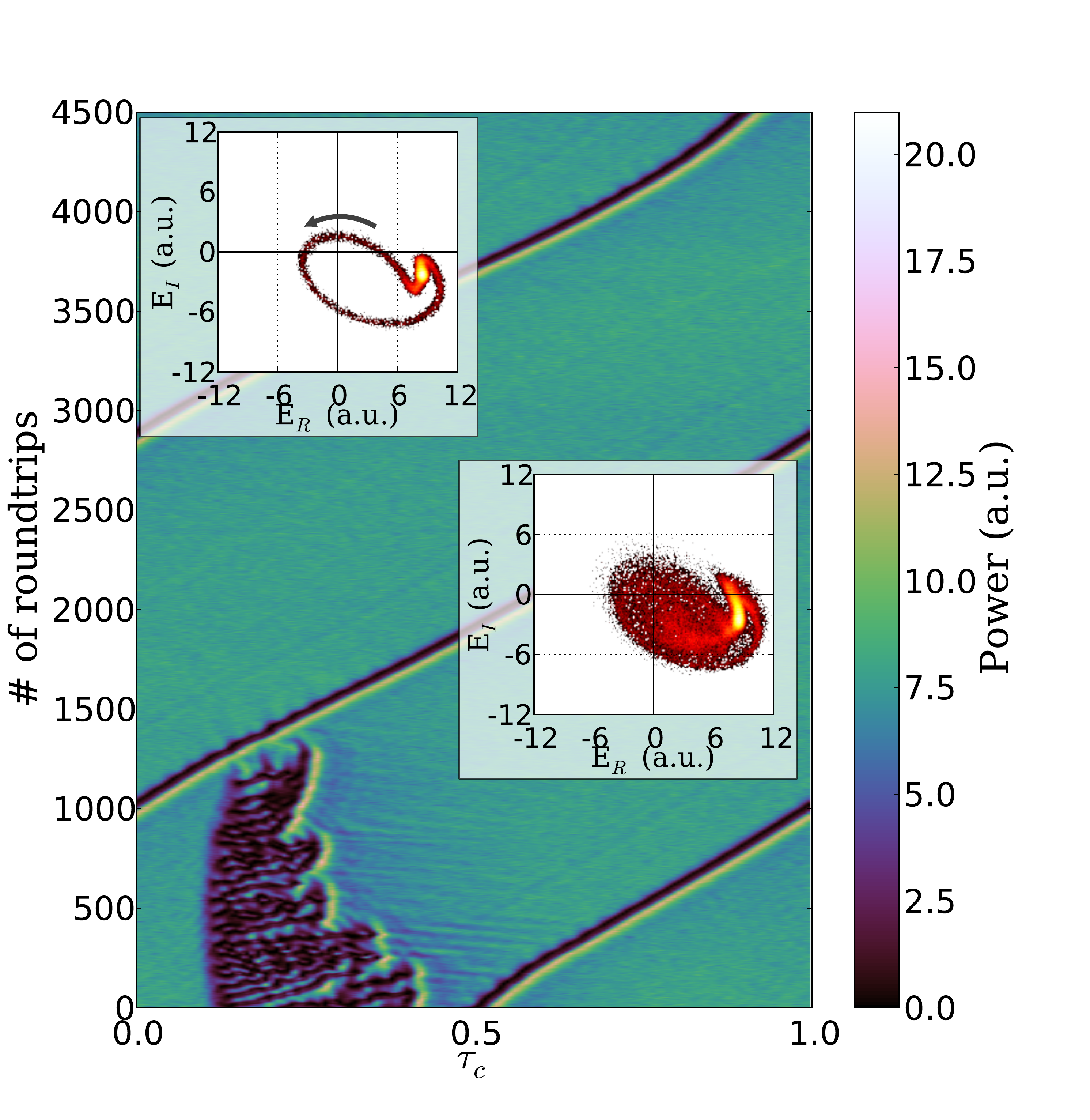}
\hspace{-1.2cm}
\includegraphics[width=4.5cm]{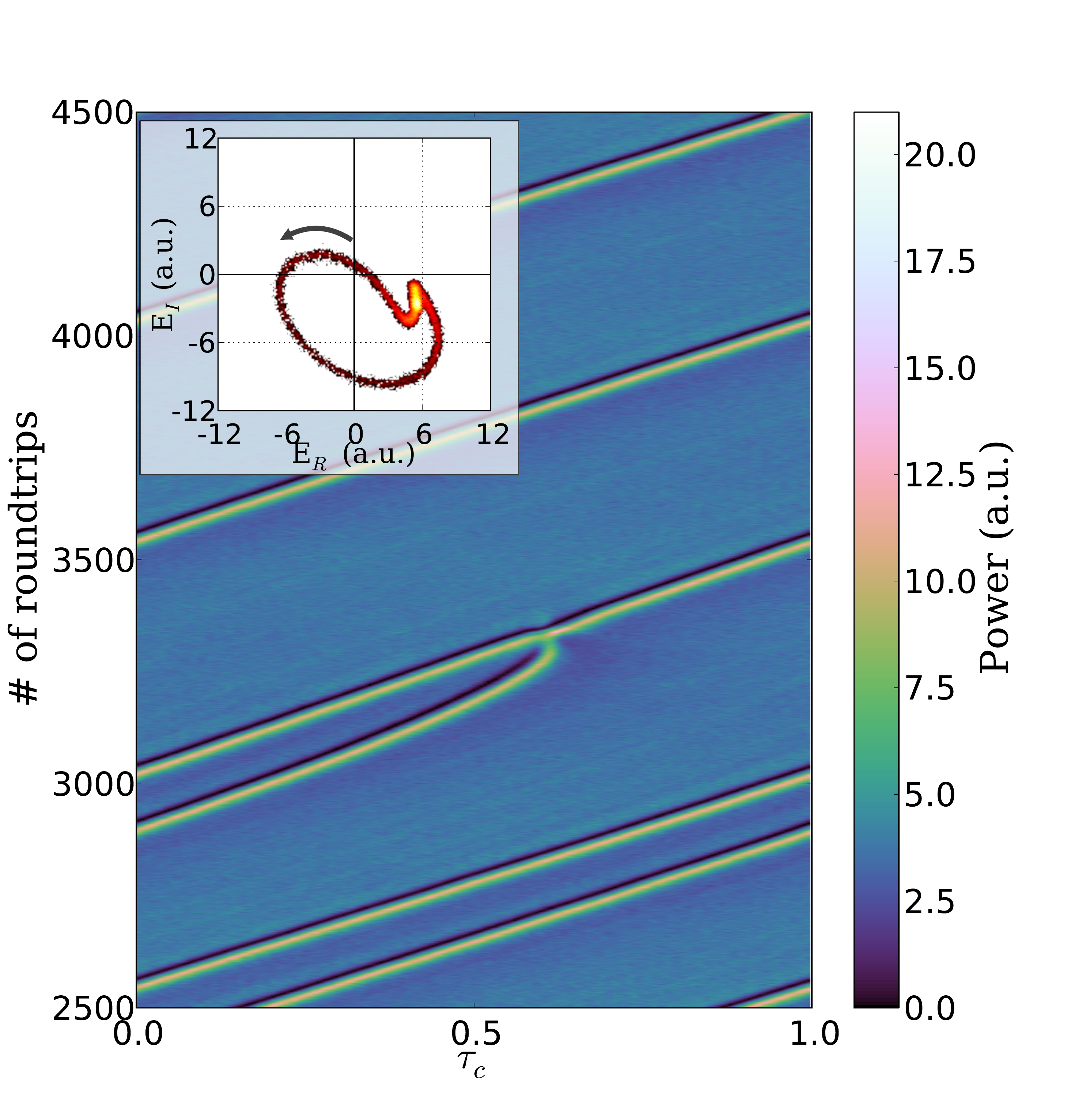}
\caption{Left: Stable phase soliton hosting a $2 \pi$ phase slip (see inset) at steady state after interaction with a turbulent state. Right: Collision and fusion of two solitons. The arrows indicate the direction of rotation along the orbit.} \label{fig:collisions}
\end{figure}
Further information about the nature of these isolated structures 
can be gained by measuring the dynamics of the optical phase. 
This is achieved by measuring simultaneously the result 
of interference between the forcing beam and the output beam of 
the system with three different dephasing conditions via a 
$3\times3$ fiber coupler, a simple formula allowing us to compute 
instantaneous intensity and relative phase \cite{Kelleher2010}.
 This measurement is shown in Fig.~\ref{fig:collisions}. 
On the left, the space-time diagram shows the coexistence of an 
isolated structure with a chaotic domain, which does not survive 
the collision with the former. The field dynamics is represented 
in the insets in the $(\mathrm{Re}(E)$,\,$\mathrm{Im}(E))$ plane. 
During the first $1000$ round-trips (lower inset), the regions 
visited by the system cover a large part of the plane, indicating 
that the dynamics cannot be well described on this 
bidimensional phase space. On the contrary, at the end of the space-time 
diagram (top inset) most of the system is locked to the external 
forcing and locally circles around the 
origin, accumulating $2\pi$ in the field phase. Localized Bloch domains carry such a chiral charge but in the present case only one stable stationary state exists and the localized states take the form of a single phase kink \cite{coullet2002localized}. As expected \cite{argentina1997colliding}, the dynamics of the field in presence of solitons projects very neatly on a bidimensional space. 

Contrary to the limit case of a pure phase dynamics described by the overdamped Sine-Gordon equation \cite{315223}, the phase rotation is accompanied here by intensity dynamics, which takes the trajectory of the system very close to the origin, where the chiral charge can disappear via the occurrence of a defect \cite{chate199917}. This provides an interpretation for the right panel of Fig.~\ref{fig:collisions}. Two identical phase solitons initially propagate at a fixed distance but at round-trip $3200$, one of them changes speed and approaches the other until they collide. This interaction does not lead to a single kink with a $4\pi$ phase rotation preserving the total chiral charge. Rather, we observe that a single phase soliton with $2\pi$ charge emerges from the collision.

\begin{figure}[t]
\includegraphics[width=0.35\textwidth]{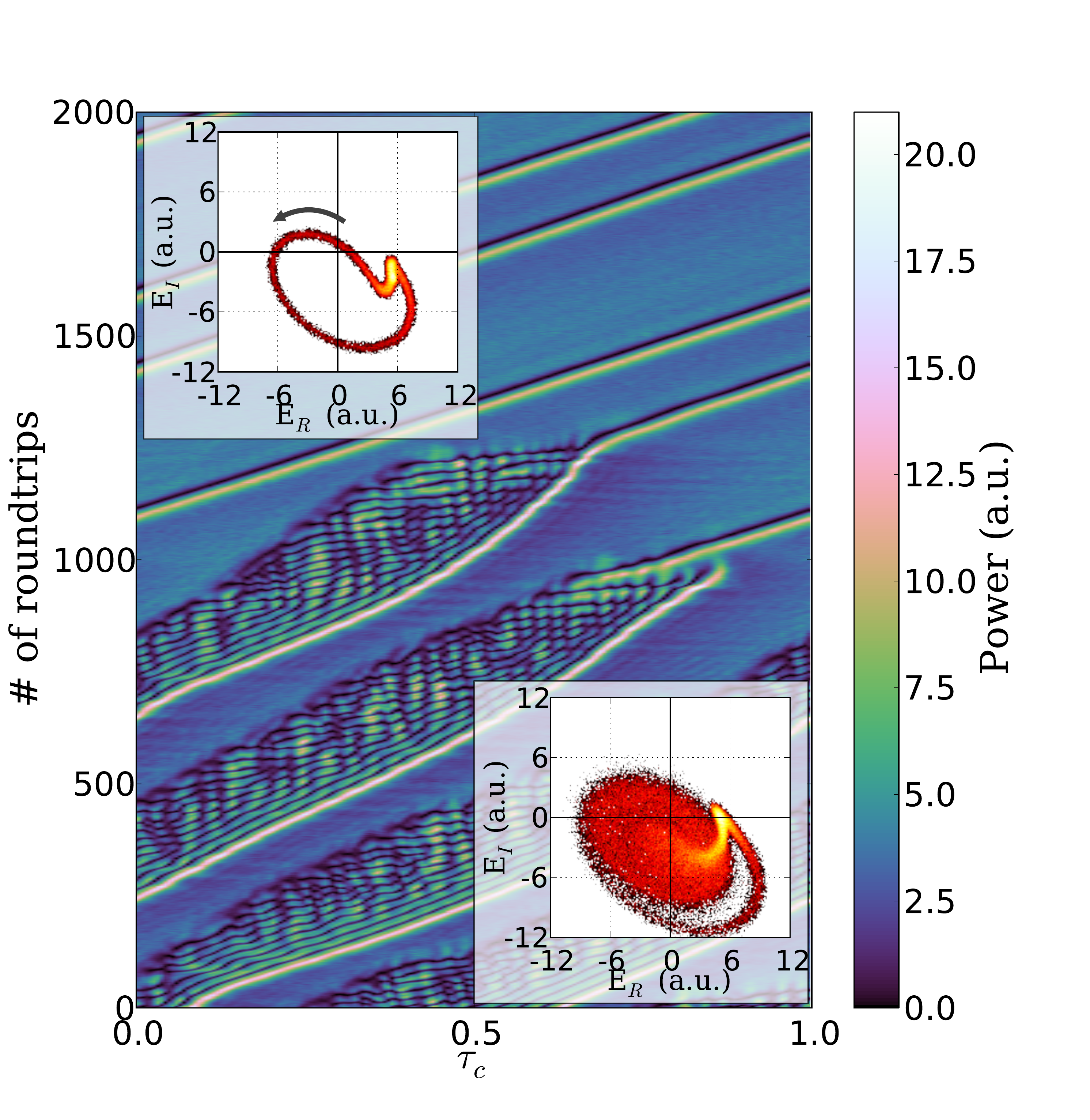}
\caption{Two phase solitons emerge from turbulent regions. Each of them hosts a $2\pi$ charge, which is already present in the preexisting chaotic domains.}
\label{fig:emerging}
\end{figure}

Fig.~\ref{fig:emerging} shows the spontaneous nucleation of a phase soliton from a turbulent state. Although the phasor plot in the lower inset suggests the presence of multiple trajectories circling around the origin, only the $2\pi$ kink at the turbulent domain border survives at the end giving rise to a phase soliton. This is actually the typical way in which phase solitons appear in our experiment. The space-time diagram of Figs.~\ref{fig:emerging} and \ref{fig:setup} are strongly reminiscent the kink breeding regime documented in \cite{chate199917}. Yet we note that the boundaries of the chaotic domains are always markedly different, which we relate to the lack of parity symmetry in the spatial dimension.

To interpret the experimental results, we extend the semiconductor models of \cite{prati2007long, prati2010static} to include the field longitudinal propagation in a coherently driven, unidirectional cavity, partially filled by an active medium. The phenomenological model for the semiconductor microscopic susceptibility allows for the dependence on frequency and carrier density of the refractive index and the gain line.

Compared to the Haus model for pulse propagation in a laser with saturable absorber \cite{haus1975theory}, our approach provides a more realistic description of the nonlinear radiation-matter interaction in the semiconductor material. Thanks to the introduction of material polarization, we do not need to introduce an \textit{ad hoc} spectral filtering section to account for material gain dispersion and avoid pulses collapse \cite{hachair2006cavity} as happens in lumped-element methods \cite{vladimirov2005model}.

The model, including the low transmission limit and neglecting diffraction, yields a set of equations for the normalized  polarization $P$ and carrier density $D$ both evolving on a faster time scale than the field envelope $E$, which read
 \begin{eqnarray}
\partial_\tau E &=&- \eta_0\partial_\eta E + \sigma \left[y-(1+i\theta)E+P\right], \label{E}\\
\partial_\tau P &=& \left[\Gamma(D)(1-i \alpha)+2 i \delta(D)\right]\left[(1-i\alpha)E D-P \right],   \label{P} \\
\partial_\tau D &=& b\left [\mu-D-\left(E^{*}P+ E P^{*} \right)/2\right], \label{D}
\end{eqnarray}
with periodic boundary conditions $E(0,\tau)=E(1,\tau)$.
\begin{figure}[t]
\includegraphics[width=0.4\textwidth]{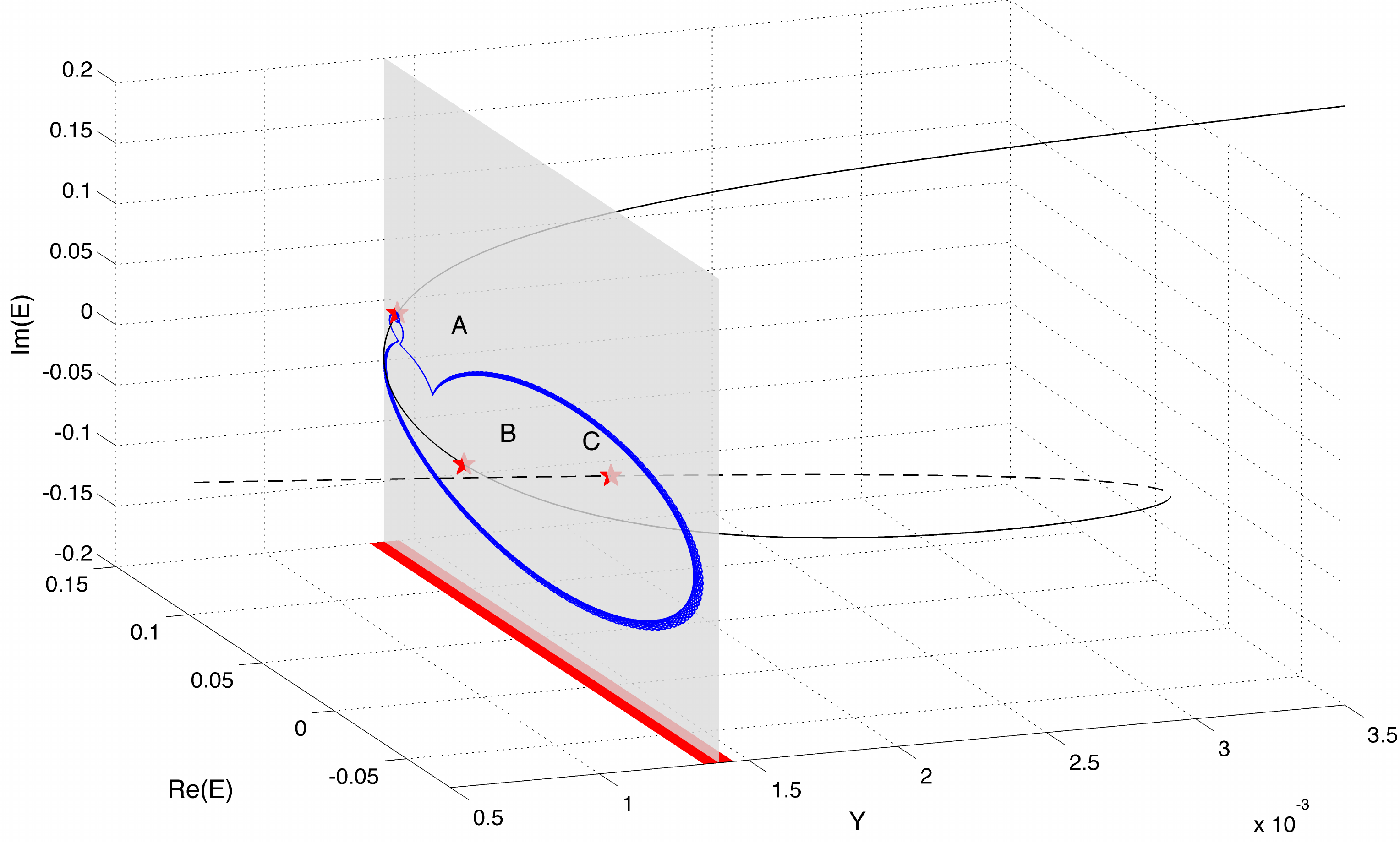}
\caption{Homogeneous synchronous states curve for $\mu=1.01$, $\alpha=3$, $\theta=-2.97$. The blue trajectory corresponds to the phase soliton numerically excited for $y=0.0014$ as in Fig.~\ref{fig:num}. The points denoted $A$, $B$ and $C$ represent a stable node, an unstable saddle, and an unstable focus, respectively.}
\label{fig:ss}
\end{figure}
The spatial variable is scaled to the length of the active medium, i.e. $\eta=z/l$, and the temporal variable $\tau$ is actually the retarded time $t+z(\Lambda-l)/(cl)$ scaled to the polarization decay time $\tau_{d}$, the parameter $\eta_0=c\tau_d/\Lambda$ is the ratio of the cavity free-spectral range to the gain linewidth, which is very small for our long cavity, meaning that even close to threshold a large number of longitudinal modes may experience gain.   
The normalized decay rates $\sigma$ and $b$ are the polarization--to--photon and polarization--to--carrier lifetime ratios, respectively; $y$ is the amplitude of the injected field (assumed real), $\alpha$ is Henry factor, $\mu$ is the scaled pump parameter. $\mu=1$ is the free running laser threshold and the value we use ($\mu=1.01$, not critical) has been chosen as the one leading to time traces most similar to those of the experiment. $\theta$ is the detuning between the frequency of the injected field (taken as the reference frequency) and the closest ring cavity resonance, multiplied by the photon lifetime.

The functions $\Gamma(D)=0.276+1.016\,D, \quad \delta(D)=-0.169+0.216\,D$ account for the finite gain linewidth and for the frequency detuning between gain and cavity, both dependent on the carrier density $D$ and were parametrized in previous work by fitting the microscopic susceptibility of the active medium \cite{prati2007long, prati2010static}. Equations~(\ref{E})-(\ref{D}) admit the stationary solution $E=|E_s|\ \mathrm{e}^{i\phi}$, $P=P_s$, $D=D_s$, with $D_s=\mu/(1+X)$, $P_s=(1-i\alpha)E_sD_s$, 

 \begin{eqnarray}
\phi &=& \arctan\left[(\alpha D_s + \theta)/(D_s-1)\right]\,, \\
y^2  &=& X\left[\left(1-D_s\right)^{2}+\left(\theta +\alpha D_s\right)^{2}\right]\,, \label{staz_curve}
\end{eqnarray}
and $X=|E_{s}|^2$. The input-output relation of Eq.~(\ref{staz_curve}) depends on the parameter $\mu$, $\alpha$, $\theta$ and it can be $S$-shaped (three-positive roots, A, B and C in Fig.~\ref{fig:ss} for $y=0.0014$), which turns out to be a necessary condition for the existence of phase solitons \cite{315223,chate199917}. In particular, the three fixed points in the subspace $(\mathrm{Re}(E),\,\mathrm{Im}(E))$ are, respectively, a stable node (A), a saddle (B) and an unstable focus (C). The corresponding phase portrait makes the system excitable in absence of propagation ($\partial/\partial z=0$) as A and B approach each other \cite{315223, chate199917,argentina1997colliding}. Linear stability analysis of the input-output relation shows that for this parameter set the phase locked solution is stable on the whole upper branch, wich disappears via a saddle-node bifurcation at the turning point, the lower branch and the negative slope branch being always unstable. The shaded region in $y$ on Fig.~\ref{fig:ss} corresponds to the stability domain of phase solitons, which grows very fast with $\mu$. Below the shaded region the locked solution becomes unstable and the system develops an irregular spatiotemporal behavior, while above the system uniformly locks to the forcing, preventing the existence of phase solitons.

\begin{figure}[t]
\includegraphics[width=4.5cm]{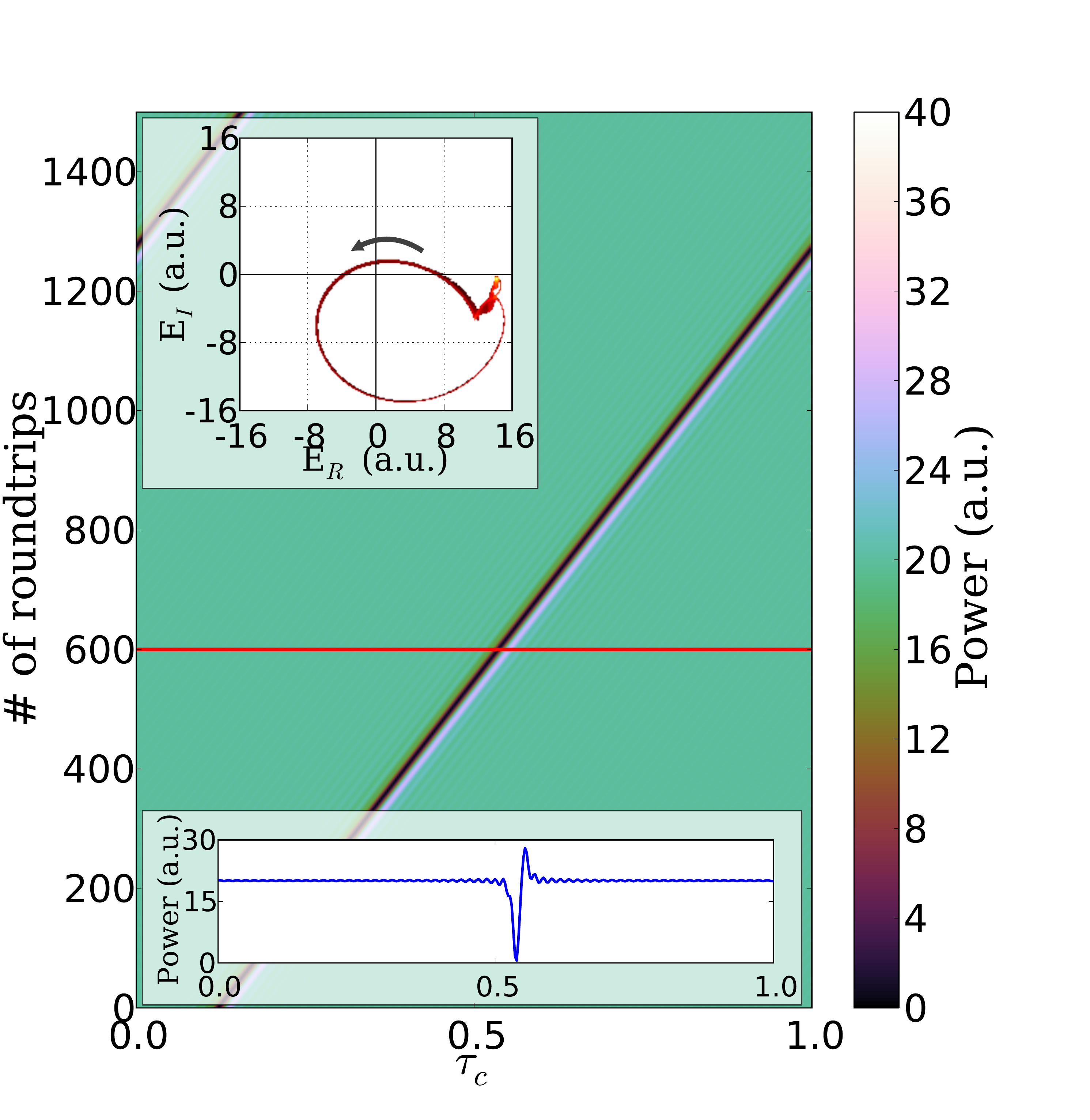}
\hspace{-1.2cm}
\includegraphics[width=4.5cm]{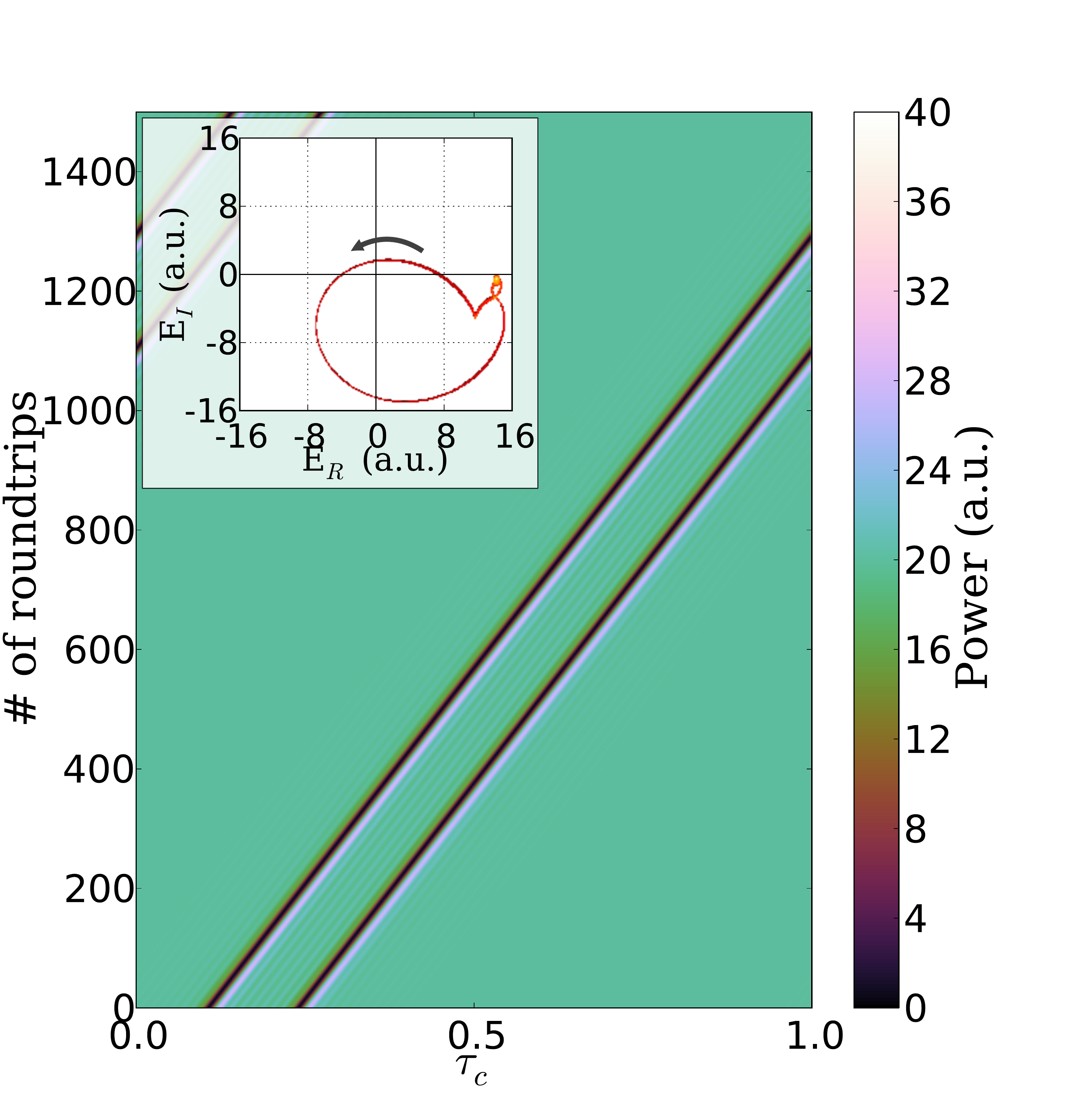}
\caption{$\sigma=3 \cdot 10^{-6}$, $b=5 \cdot 10^{-4}$,$y=0.0014$.  The other parameters are as in Fig.~\ref{fig:ss}. Left: Stable phase soliton seeded by a $2\pi$ phase
slip superimposed to the homogeneous synchronous state $A$. Bottom left: field intensity profile within one roundtrip. Right: Two coexisting phase solitons at steady state which display the same phase portrait (insets)}
\label{fig:num}
\end{figure}
In the numerical simulations a stable phase soliton can be obtained with initial conditions corresponding to the locked state $A$, to which we superimpose a positive phase kink of $2\pi$ along $z$ for the field $E$. The final state is a traveling pulse with a $2\pi$ chiral charge and a width of few hundreds of ps, which reproduces well the experimental results (Fig.~\ref{fig:num}, left). Its phasor representation in the subspace $(\mathrm{Re}(E),\,\mathrm{Im}(E))$ is shown in Fig.~\ref{fig:ss} by a blue line lying in the shaded plane, which confirms that the dynamics can be effectively embedded in the 2D subspace. 

Multiple structures separated by varying distances can also exist simultaneously (Fig.~\ref{fig:num}, right). Within parametric regimes compatible with the experiment, any structure missing the chiral charge or with the wrong charge sign is invariably unstable, though long transients can occur.

It has been shown \cite{315223} that a forced two--level laser model (Maxwell-Bloch equations) close to threshold can be reduced to a modified Ginzburg--Landau equation which admits a free energy equivalent to the potential of the Frenkel-Kontorova model \cite{PhysRevLett.56.724,315223,braun2004frenkel}. Under certain parametric conditions the laser dynamics turns out to be slaved to the phase, which obeys an overdamped sine--Gordon equation \cite{315223}. Although the latter limit does not apply to our experiment, Eqs. (\ref{E}--\ref{D}) can be reduced to a modified forced Ginzburg--Landau equation using experimentally meaningful approximations. Precisely, we assume that the laser is very close to threshold setting $\mu-1=\epsilon\ll 1$, and that the injected amplitude $y$ and the detuning between the free running laser frequency and the injected frequency, which in our model is $\theta+\alpha$, are both of order $\epsilon$.
A multiple scale analysis truncated at order $\epsilon^2$ has allowed us to eliminate the polarization and carrier density while preserving the main features of the semiconductor susceptibility (finite width and position of the gain peak). The final equation for the field envelope $E$ reads

\begin{eqnarray}
\eta_0\partial_\eta E+\partial_\tau E&=&\sigma \left\{y+\left[\mu-1-i(\mu\alpha+\theta)\right]E\right.\nonumber\\
&&-(1-i\alpha)|E|^2E \nonumber\\
&& +\left. d\left[\eta_0^2\partial^2_\eta-2i\delta(1)\eta_0\partial_\eta\right] E\right\}\,, \label{GL}
\end{eqnarray}
with $d=\left[\Gamma(1)^2(1-i\alpha)\right]^{-1}$. The equation differs from that of \cite{315223} mainly for the last term, which describes a nonsymmetric gain line, and for the presence of the Henry factor $\alpha$ which implies that the medium is dispersive even at resonance, two typical features of semiconductors. 
In a framework moving at velocity $\eta_0$ (i.e. co-moving with the soliton), Eq.~(\ref{GL}) lacks the parity symmetry $\eta\rightarrow-\eta$ of the standard Ginzburg--Landau equation due to non-vanishing $\delta(1)$. However, unless that parameter is set to an unrealistically large value, Eq.~(\ref{GL}) supports both chiral charges contrary to the experiment and the complete model. Therefore, we conclude that the main mechanism for chiral selection lies in the non-instantaneous medium dynamics, which heavily breaks the $\eta\rightarrow-\eta$ symmetry in a propagative system. 

In conclusion, we have observed robust solitary optical structures which propagate in a highly multimode semiconductor ring laser with coherent forcing. We have measured directly their optical phase and showed that they carry a chiral charge whose sign is decisive for their stability. Numerical simulations of a physical model taking into account both the semiconductor material susceptibility and the geometry of the experiment allow the interpretation of the experimental observations in terms of (multistable) phase solitons. In addition, multiple scale analysis has allowed to understand these solitons as the elementary excitations of oscillatory media under nearly resonant forcing close to a commensurate-incommensurate transition. Finally, downscaling the whole experiment to monolithic semiconductor ring lasers may lead to robust and ultrafast phase solitons for phase information encoding and regeneration in coherent optical communications \cite{slavik2010all,radic2010optical}.

\begin{acknowledgments}
The authors would like to thank Dr. Lionel Gil for many helpful discussions. BK acknowledges support from the Irish Research Council. This work was conducted in part under the framework of the INSPIRE programme, funded by the Irish Government's Programme for Research in Third Level Institutions, Cycle 4, National Development Plan 2007-2013, supported by the European Union Structural Fund. LC acknowledges financial support from the MIUR project number PON02-0576. FG, GT and SB acknowledge funding from Agence Nationale de la Recherche through grant number ANR-12-JS04-0002-01. BK and SB acknowledge funding from Campus France through grant number 29872PF.

\end{acknowledgments}

\appendix

\bibliography{phasesol}

\begin{thebibliography}{40}
\bibitem[{gre()}]{greenfootnote}
\bibinfo{note}{The cubehelix colorscheme is designed to be monotonically
  increasing in terms of its perceived brightness both in color and on black
  and white prints.}
\end{thebibliography}

\end{document}